# A COMPARATIVE STUDY OF AGGREGATE TCP RETRANSMISSION RATES

K. Pentikousis[*], H. Badr[**] and A. Andrade[***]

[*] VTT Technical Research Centre of Finland, Kaitoväylä 1, FI-90571 Oulu, FINLAND

[**] Department of Computer Science, Stony Brook University, Stony Brook, NY 11794-4400, USA

[***] Yahoo Inc., Sunnyvale, CA 94089, USA

email: [*] kostas.pentikousis@vtt.fi , [**] badr@cs.stonybrook.edu , [***] aandrade@yahoo-inc.com

**Abstract** — Segment retransmissions are an essential tool in assuring reliable end-to-end communication in the Internet. Their crucial role in TCP design and operation has been studied extensively, in particular with respect to identifying non-conformant, buggy, or underperforming behaviour. However, TCP segment retransmissions are often overlooked when examining and analyzing large traffic traces. In fact, some have come to believe that retransmissions are a rare oddity, characteristically associated with faulty network paths, which, typically, tend to disappear as networking technology advances and link capacities grow. We find that this may be far from the reality experienced by TCP flows. We quantify aggregate TCP segment retransmission rates using publicly available network traces from six passive monitoring points attached to the egress gateways at large sites. In virtually half of the traces examined we observed aggregate TCP retransmission rates exceeding 1%, and of these, about half again had retransmission rates exceeding 2%. Even for sites with low utilization and high capacity gateway links, retransmission rates of 1%, and sometimes higher, were not uncommon. Our results complement, extend and bring up to date partial and in-complete results in previous work, and show that TCP retransmissions continue to constitute a non-negligible percentage of the overall traffic, despite significant advances across the board in telecommunications technologies and network protocols. The results presented are pertinent to end-to-end protocol designers and evaluators as they provide a range of "realistic" scenarios under which, and a "marker" against which, simulation studies can be configured and calibrated, and future protocols evaluated.

## 1. Introduction

Measuring end-to-end path characteristics is a very active network research field. Passive and active measurement methodologies are employed, in particular, to study the behaviour of the Transmission Control Protocol (TCP) and estimate several performance gauges; for example, packet loss and round-trip times, to name a couple. Most studies focus on individual, per connection metrics and often ignore aggregate behaviour. This article centres on segment retransmission rates in aggregate TCP traffic as observed at the egress/ingress points of six different large sites, including three university campuses (stub autonomous systems), two traffic aggregation sites, and a large research centre. TCP retransmission and packet loss estimation measurements have long been a study item as they are a fundamental component of TCP's reliable byte-stream service and are tightly coupled with error recovery and congestion avoidance actions. We complement and extend related work (discussed in Section 4) by quantifying the range of this important gauge of overall TCP performance, based on data derived from network traces made available by the National Laboratory of Applied Network Research Passive Measurement and Analysis (NLANR/PMA) [1]. This repository is currently maintained by CAIDA (www.caida.org). Although researchers may be able to "extract" TCP retransmission rates from other studies, to the best of our knowledge there is no adequate, readily available, detailed study reporting on recent data for real-world TCP retransmission rates that takes into consideration traces from several passive monitoring sites in a comparative and consistent manner. While some studies report on average packet loss rates across different protocols, typically in large backbones, TCP-specific studies tend to focus on single flow, typically active, measurements. Our attempt at estimating the level of TCP retransmission takes a different approach. Specifically, we report on the distribution of segment retransmission rates in aggregate TCP traffic as seen, for example, at the egress points of large stub autonomous systems and traffic aggregation sites.

Besides providing an important record for future reference, this paper contributes to the discussion about future Internet protocol design. In short, this study provides "hard", timely and reliable quantitative data about TCP behaviour in the aggregate, as measured by the experiences of real-world flows from different networks. It provides a "marker" against which the performance of future protocols may be evaluated. Moreover, we present the results in a comparative manner, which allows drawing a fuller picture about what is currently happening with respect to TCP retransmissions.

This paper is organized as follows. In Section 2, we present the methodology employed in this study. In Section 3 we present our results and in Section 4 we compare them with previous work. In Section 5 we discuss the importance of the measurement results and how they relate to future Internet protocol design. Finally, we conclude this paper in Section 6 outlining future work items.

## 2. Methodology

NLANR/PMA traces are captured at a variety of large sites, including university campuses. A passive monitoring point (PMP) is typically attached to an egress/ingress gateway at a given site and records each packet passing by, as illustrated in Figure 1. Traffic originates from (or is destined to) a wide variety of hosts ranging from super computers connected via direct links to the campus backbone, to laptops and other mobile devices connected via wireless local area networks (WLANs) to the Internet. The traffic traces are captured and stored in time sequenced headers (TSH) format and contain the multiplexed traffic from (to) the different traffic source (destination) points of the site, recorded in each direction separately. A single TSH record is 44 bytes long, comprising: (a) time and interface number; (b) the standard IP packet header (IP options are not recorded); (c) the first 16 bytes of the IP packet payload, which, for a TCP segment payload, constitute the first 16 bytes of the standard TCP header. The TCP checksum, urgent pointer, and TCP options (if any) are not included in a TSH record.

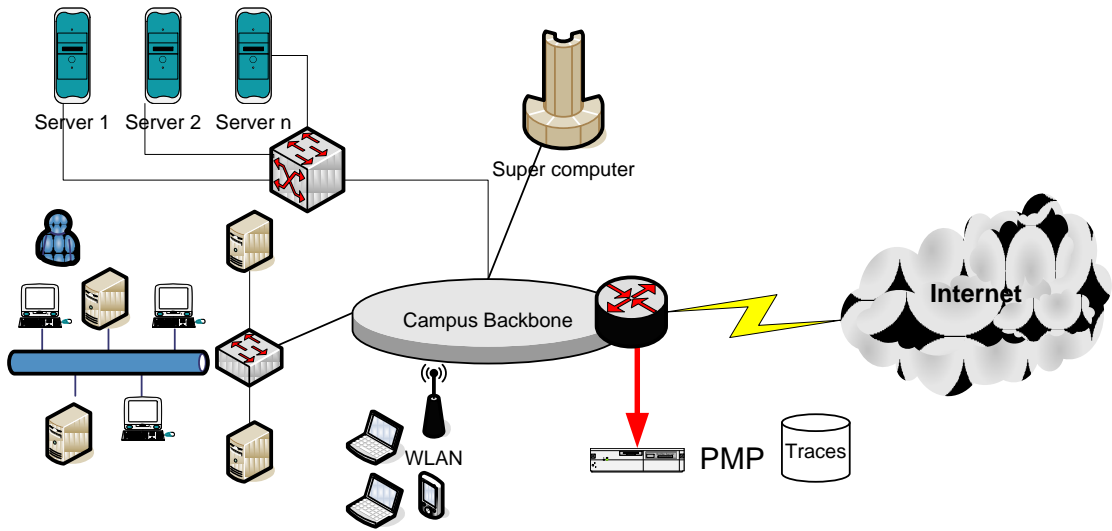

Figure 1. Passive network traffic monitoring

Using software we developed [2], we reconstruct the multiplexed TCP stream in each trace and identify retransmitted segments. We define the retransmission rate $\eta = \dfrac{R}{N}$, where $N$ is the number of TCP data-carrying segments in a single trace, and $R$ the number of retransmitted segments amongst the $N$. Complementing the results discussed in Section 4 below on related work, we focus on quantifying the characteristics of $\eta$ across several sites in a comparable manner. Note that $\eta$ is a lower bound on the proportion of retransmissions along the full end-to-end paths, since seg-segments lost in the network before the PMP and subsequently retransmitted are not accounted for, as is the case in related previous work, such as [3][4][5], as well.

## 3. Results

We processed 1337 traces, cumulatively covering 33 hours and containing 731 GB of IP traffic, 91% of which is TCP data-carrying segments ($642 \times 10^6$ segments – see Table 1). The traces were captured at six PMPs located in the United States, as summarized in Table2. Three PMPs are attached to gateway routers of universities (Columbia University–BWY, Colorado State University–COS, and Old Dominion University–ODU); one is located at a research centre (Pittsburgh Supercomputing Center–PSC); and two are at traffic aggregation points (AMPATH–AMP, located at the NAP of the Americas in Miami, Florida; and the Front Range GigaPOP–FRG, located in Denver, Colorado). Traces are captured at varying times of the day/month/year, each trace corresponding to the unidirectional traffic (incoming or outgoing) passing by the PMP interface during 90 seconds. The results presented in this paper correspond to traces dating mostly from 2004-06 (72.9%), with some from 2001-03 (9.8%), while 17.3% of the traces had no specific date information but were captured between 2001 and 2006. The cumulative proportion of TCP data-carrying segments across all 1337 traces which are retransmissions is 2.4% (Table 1). The complete list of traces is available from the authors.

Table 1
Cumulative trace summary

| | |
|---|---|
| Number of Traces | 1337 |
| Cumulative Duration (hours) | 33.4 |
| Total IP Bytes ($\times 2^{30}$) | 731.072 |
| — Percentage (%) of TCP traffic | 91.1% |
| Total TCP data segments | 641799931 |
| — Percentage (%) of retransmitted segments | 2.4% |

## 3.1 Aggregate TCP Retransmission Distribution

For an interface (incoming or outgoing) of a given PMP with, say, $T$ traces, $\bar{\eta}$ denotes the average of the $T$ values for $\eta$ obtained from these traces: $\bar{\eta} = \frac{1}{T}\sum_{t=1}^{T}\eta_t = \frac{1}{T}\sum_{t=1}^{T}\frac{R_t}{N_t}$. On the other hand, $\hat{\eta}$ denotes the overall retransmission rate for the $T$ traces: $\hat{\eta} = \frac{\sum_{t=1}^{T}R_t}{\sum_{t=1}^{T}N_t}$. As shown in Table 2, $\bar{\eta}$ ranges between 0.48% and 3.6%; and $\hat{\eta}$ between 0.45% and 3.38%, except for the second ODU interface, ODU.2, which has $\hat{\eta}$ = 7.64%. The trace sets for AMP, BWY, COS, and PSC (858 traces in total) do not contain any traces with segment retransmission rate $\eta$ > 5.14% (see Table 3). The FRG and ODU sets (479 traces), on the other hand, have a total of 26 traces with 10% < $\eta$ < 32.5%; ODU.2 has a further three traces with quite extraordinary $\eta$ values of 62.81%, 63.94%, and 75.84%, which significantly distort its $\hat{\eta}$.

Table 2
Trace/site and monitored interface summary

|  | AMPATH | | Pittsburgh Super-computing Center | | Front Range Giga-POP | | Columbia University | | Colorado State University | | Old Dominion University | |
|---|---|---|---|---|---|---|---|---|---|---|---|---|
| Capacity (Mb/s) | 622.08 | | 2488.00 | | 622.08 | | 155.52 | | 155.52 | | 155.52 | |
|  | AMP.1 | AMP.2 | PSC.1 | PSC.2 | FRG.1 | FRG.2 | BWY.1 | BWY.2 | COS.1 | COS.2 | ODU.1 | ODU.2 |
| No. of traces | 101 | 101 | 102 | 102 | 119 | 119 | 115 | 113 | 113 | 111 | 123 | 118 |
| Cum. duration (min) | 151 | 151 | 152 | 152 | 178 | 178 | 174 | 171 | 170 | 167 | 183 | 176 |
| Overall link utilization (%) | 2.6 | 3.4 | 2.7 | 4.1 | 20.0 | 18.8 | 14.5 | 9.9 | 20.6 | 23.3 | 24.6 | 19.3 |
| Bytes ($10^6$) | | | | | | | | | | | | |
| IP | 18103 | 23969 | 78068 | 115420 | 165819 | 156058 | 29493 | 19665 | 40839 | 45458 | 52534 | 39557 |
| TCP | 14878 | 21730 | 69173 | 106100 | 158432 | 149181 | 16518 | 17462 | 37760 | 41737 | 47800 | 34122 |
| TCP segments ($10^3$) | | | | | | | | | | | | |
| (a) total) | 12956 | 17850 | 58293 | 85092 | 132922 | 141000 | 17053 | 17431 | 35616 | 47198 | 44258 | 32131 |
| (b) retransmitted | 79 | 175 | 372 | 381 | 3156 | 4772 | 233 | 236 | 968 | 990 | 1504 | 2454 |
| $\hat{\eta} = (b) \div (a)$ (%) | 0.61 | 0.95 | 0.64 | 0.45 | 2.37 | 3.38 | 1.36 | 1.35 | 2.72 | 2.10 | 3.40 | 7.64 |
| $\bar{\eta}$ (%) | 0.65 | 0.95 | 0.64 | 0.48 | 1.86 | 2.73 | 1.30 | 1.26 | 2.50 | 1.91 | 3.33 | 3.66 |

Figure 2 presents box plots for the six PMPs, on a per interface basis. Each box plot shows the distribution of $\eta$ values for the individual traces of the interface. The plots are grouped into two sets: the sites with relatively high-bandwidth OC-12 and OC-48 links (Figure 2(a)), and the three university sites with lower-bandwidth OC-3 links (Figure 2(b)); note that the two sets are not displayed to the same vertical scale.

AMP and PSC traces (Figure 2(a)), collected from OC-12 and OC-48 interfaces, respectively, and with mostly very low utilizations averaging 2% to 4% (Table 2), have $\tilde{\eta} \cong 0.5\%$ (except for AMP.2 for which $\tilde{\eta}$ is still below 1%), where $\tilde{\eta}$ denotes the median. These retransmissions are presumably mostly due to losses or other network pathologies [6][7]. FRG traces (Figure 2(a)) come from O-12 interfaces with relatively high utilizations averaging almost 20% (Table 2), but nevertheless also have $\tilde{\eta} \cong 0.5\%$; this, however, does not tell the whole story, as we shall see when we present Table 3 below.

In contrast, traces from the two university campuses COS and ODU (Figure 2(b)) have median values $\tilde{\eta}$ mostly above 2%, and may even have third quartile values $Q_3 > 3\%$ (COS.1 and ODU.1). These traces come from access links with the lowest bandwidth (OC3), operated at higher utilization rates (Table 2); indeed, each interface for COS and ODU has ≥ 20% of its traces with a link utilization exceeding 30%. They are therefore, presumably, more likely to experience heavier congestion in their own right, due to the bursty nature of TCP traffic, causing segment retransmission over and above that due to losses occurring beyond them, in the Internet at large. On the other hand, BWY (Figure 2(b)) is also a university campus with an OC-3 link, but has lower link utilizations (Table 2) with consequently lower median and third quartile values: $\tilde{\eta} \cong 1\%$ and $Q_3 < 2\%$.

Ten of the twelve interfaces have $\tilde{\eta} > \bar{\eta}$ (see Figure 2 and Table 2), indicating a tendency – particularly sharp in the cases of FRG.1/2 and ODU.1 interfaces – to skew to the right (i.e., values of $\eta$ that are greater than the mean $\bar{\eta}$ occur more frequently than values $\eta < \bar{\eta}$). The only exception to this are the two COS interfaces, for which the means are virtually equal to the medians (*i.e.*, $\eta < \bar{\eta}$ and $\eta > \bar{\eta}$ are equally likely).

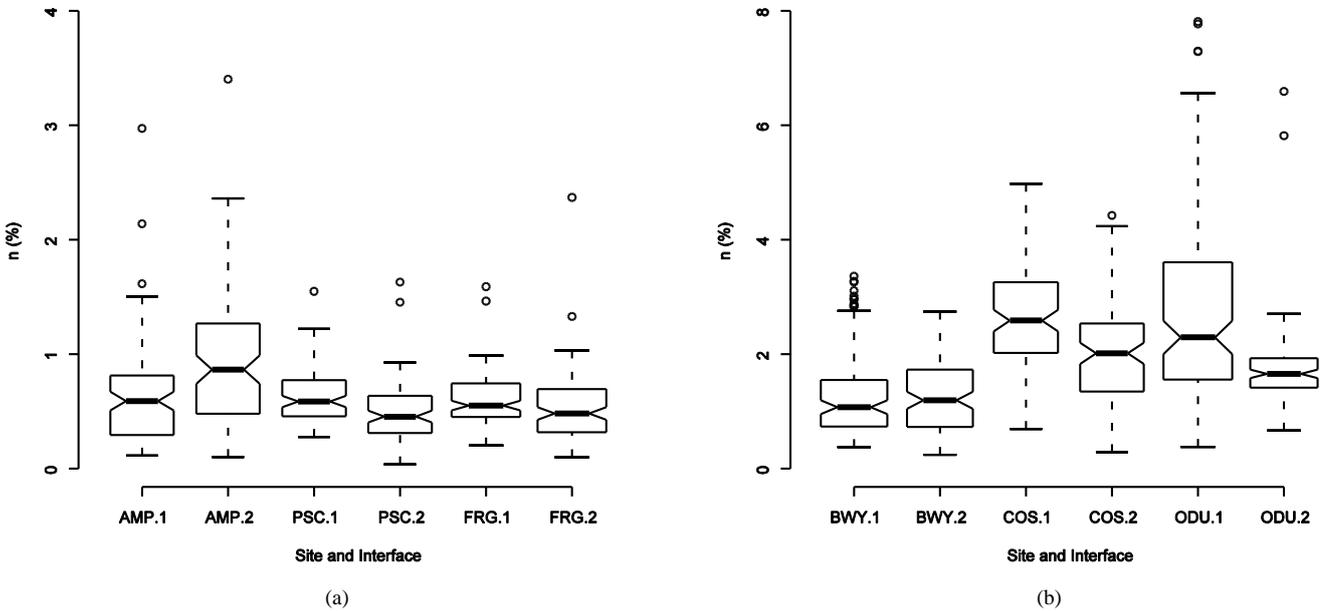

Figure 2. Box plots of TCP-segment retransmission-rate percentages. A box contains the middle 50% of the data set range; the line in the middle represents the median, $\tilde{\eta}$; and the "hinges" the $Q_1$ and $Q_3$ quartiles. The horizontal line below the box demarcates the smallest value; the one above extends to $\min\{\eta_{max}, 1.5 \times (Q_3 - Q_1)\}$. Values outside this upper "whisker" line, shown as circles, are considered outliers. Mean values, $\bar{\eta}$, are given in Table 2.

The box plots of Figure 2 do not give a sufficiently detailed picture of the tails of the distributions for $\eta$. These are presented in **Table** 3. Note the relatively long tails of the distributions for FRG's high-utilization OC-12 interfaces, as compared to AMP and PSC (and to the OC-3 BWY interfaces as well, which have lower overall utilizations than FRG – see **Table**2). The OC-3 interfaces for COS and ODU have higher overall utilizations that are comparable to each other (and, to a lesser extent, to FRG as well) in their details, yet display distinctly different characteristics in the tails of their distributions.

Table 3
Tails of the distributions for $\eta$

| | AMPATH | | Pittsburgh Supercomputing Center | | Front Range GigaPOP | | Columbia University | | Colorado State University | | Old Dominion University | |
|---|---|---|---|---|---|---|---|---|---|---|---|---|
| Proportion of traces with $\eta >$ | AMP.1 | AMP.2 | PSC.1 | PSC.2 | FRG.1 | FRG.2 | BWY.1 | BWY.2 | COS.1 | COS.2 | ODU.1 | ODU.2 |
| 1% | 9.9% | 39.6% | 10.8% | 2.0% | 9.2% | 13.5% | 55.7% | 59.3% | 91.2% | 79.3% | 95.1% | 94.9% |
| 2% | 3.0% | 6.9% | | | 7.6% | 11.8% | 15.7% | 15.0% | 71.7% | 48.7% | 59.4% | 21.2% |
| 3% | 1.0% | 1.0% | | | 7.6% | 10.9% | 4.4% | | 31.9% | 15.3% | 34.2% | 5.9% |
| 4% | 1.0% | | | | 7.6% | 10.9% | | | 8.0% | 1.8% | 20.3% | 5.9% |
| 5% | 1.0% | | | | 7.6% | 10.9% | | | | | 13.8% | 5.9% |
| 10% | | | | | 6.7% | 9.2% | | | | | 3.3% | 4.2% |
| 15% | | | | | 5.9% | 9.2% | | | | | 1.6% | 3.4% |
| 20% | | | | | 2.5% | 6.7% | | | | | 1.6% | 3.4% |
| 25% | | | | | 0.8% | 2.5% | | | | | 1.6% | 2.5% |
| 30% | | | | | | 1.7% | | | | | 0.8% | 2.5% |

### 3.2 Aggregate TCP Retransmission Distance

RFC 3357 [8] defines loss distance for a lost packet in a given packet stream as the number of packets transmitted since the previous loss. We extend this definition to retransmissions in aggregate TCP traffic in a straightforward manner: for a retransmitted TCP segment, the retransmission distance $\nu, \nu \geq 1$, is defined as the number of TCP data-carrying segments in the multiplexed traffic since the previous retransmitted segment (note that, according to our definition, we do not count non-TCP packets). Since we are dealing with multiplexed traffic from multiple TCP sources, two successive retransmitted segments, yielding some $\nu$ value, need not belong to the same TCP connection.

We have observed that, despite a high degree of multiplexing in the traces, a very substantial proportion of retransmissions in the aggregate traffic occur with no intervening successfully-transmitted TCP segment in between ($\nu = 1$), or with at most a couple such segments (this is reminiscent of correlated retransmissions in a single TCP connection). One typical example from a PSC trace (with $\bar{\eta} = 1.08\%$) will serve to illustrate this general characteristic, which is highly significant from a TCP perspective. At the minimum value $\nu = 1$ Pr.$[\nu = 1] = 0.121$, and Pr.$[\nu \leq 3] = 0.191$. Further analysis of the traces is required to determine the underlying distributions for $\nu$ and whether $\nu$ values are correlated. It is, of course, well known that an individual sender can experience correlated losses (due to drop tail effects at network interfaces and corruption in wireless links) in its end-to-end path (see for example [9]), while its retransmission pattern depends on the TCP variant. The question here (especially as it applies, for example, to the AMP and PSC sets of traces, which have high bandwidth and low link utilizations) is whether the aggregation of a sufficiently large number of such TCP flows, each forced to retransmit due to losses at different points in the Internet in a manner conjectured to be largely independent the one from the other, might not collectively produce some sort of "random drop effect" in the multiplexed flow observed at the egress/ingress PMP of a large site. Studying and modelling this behaviour is part of our ongoing work.

## 4. Related Work

Paxson [7] performed active measurements between 35 sites, using 100-KB TCP bulk transfers, and reported several adverse network characteristics, including large packet loss rates (2.7% in December 1994; 5.2% in November-December 1995). With respect to TCP retransmissions he observed that for correct implementations only about a quarter of them were redundant. He concluded that "TCP's retransmission strategies work in a sufficiently conservative fashion", highlighting that "standard-conformant RTO calculations and deploying the SACK option together eliminate virtu-ally all of the avoidable redundant retransmissions." More recently, Allman et al. [10] used a measurement mesh network to quantify the effectiveness of counting TCP segment retransmissions as an estimator of packet loss rate. They showed that TCP retransmissions overestimate actual packet loss by more than 10% in 33% of Reno and 25% of SACK transfers, respectively, each comprising 5000 packets.

Detecting and classifying out-of-sequence segments and spurious retransmissions has received much attention as well (see, for example, [3][5][11][12] and the references therein), although such issues are beyond the scope of this paper. Nevertheless, based on [7][10], it is clear, in light of the results presented in the previous section, that packet loss rates in excess on 1.5-2% are not a rare oddity but, rather, occur quite frequently, despite the tremendous increases in network capacities in recent years.

Our results complement those of Rewaskar et al. [3], who use a passive state-machine approach to detect packet losses and classify them according to prevalent root cause. Although there are certain discrepancies in their results tables, one can "extract" $\eta$. In brief, the $\eta$ values surmised are in the range 0.66% to 1.88% in six of their data sets, and between 5.79% and 6.65% for a seventh set, yielding a weighted average of 1.4 to 1.6%. Clearly, these values, which correspond to seven single but longer traces, are within the range of our results. Jaiswal et al. [4][5] report on a total of eight long, mostly Tier-1, backbone traces dating back from 2002. Five of the eight traces, of duration 6 hours each, come from OC-12 (622 Mbps) backbones; the remaining three traces (1~2 hours each) from OC-48 (2.5 Gbps). As with [3], the focus of the work is on determining the cause of out-of-sequence segments. Values for $\eta$ range from 1.5% to around 5.3% across the eight traces, with a weighted average of about 3.5%. These values are within, but noticeably on the higher side of, our OC-12 and OC-48 results which report on a much larger set of much shorter but more recent traces totalling 16 hours. Note that the authors in [3][4][5] are unable to determine the cause of all out-of-sequence packets observed in their traces. While these "unexplained" segments do not greatly impact the potential values for $\eta$, they do make it difficult to report a categorically precise set of values in a brief, summary overview such as the one presented in this paper.

## 5. Discussion

The results presented in this paper add an important reference point to the ones discussed in the previous section. They are of particular interest to researchers studying the development of computer networks and of TCP performance and behaviour over time. Segment retransmissions are tightly coupled with congestion avoidance actions in TCP, and are a fundamental component of the reliable byte stream service that TCP provides to applications. They are also a good indicator of low performance. Retransmissions "cost" not only in the amount of extra traffic that has to be injected into the network, but also in terms of forcing a connection to remain idle occasionally. As bandwidth capacities grow, the cost of remaining idle increases.

In presenting salient characteristics of the distributions of aggregate TCP retransmission rates from six different sites we are quantifying an important metric in a comparative manner. Our measurements, based on the experiences of regular, real-world TCP flows, show that segment retransmissions continue to consume a non-negligible proportion of network capacity despite advances in networking technologies, the tremendous growth in link capacities, and the deployment, as evidenced by [13][14], of newer TCP versions such as NewReno and SACK. This contrasts with results reported from measurement infrastructures such as PingER [15], and the so-called "network weather maps" popular with Network Operation Centres, showing near-zero packet losses [16].

Finally, our results shed a different light on what constitutes a realistic retransmission environment for both simulation and analytical TCP studies. Matthews and Cottrell [15] make it clear that, in the past, every single link capacity upgrade has been quickly consumed by new increases in traffic levels. End-to-end protocols have to ensure that the complete application payload is successfully delivered to the peer and, thus, ought to be tested under different conditions. The results presented in this paper permit the researcher testing, say, the resiliency of a TCP modification, to calibrate a range of realistic, simulated, "baseline" network configurations by changing traffic load and mix, and/or modifying link capacities and queue lengths, such that existing TCP variants are made to yield retransmission characteristics presented by our results; then using the calibrated configurations to evaluate TCP variants in an effective manner. Alternatively, one could use an error model which induces such retransmission rates, while preserving the general distributional characteristics presented. This is an area of past [17] and future work for us. Employing simulations that rely exclusively on tweaking the traffic load in a more or less arbitrary manner to determine the retransmission rate may not be the best or only way forward. Retransmissions are not only caused by excessive loads. Checksum failures, acknowledgment compression, packet reordering, and spurious timeouts are all causes for retransmissions. The retransmission ranges documented in this study should be of interest to researchers working on analyzing and modelling TCP error patterns, and on developing error models for simulation studies.

## 6. Conclusion

TCP retransmission is a product of several factors (network capacities, traffic mix and load, policing, queue management, and so on) and essentially depends on implementation particulars (and bugs). We quantified aggregate TCP retransmission rates by analyzing a large set of traffic traces captured at the egress/ingress points of six large network sites, presented their distributional characteristics, and compared our results with related work. We observed aggregate TCP retransmission rates in the range 1-2% in about a quarter of the traces, and another quarter had a retransmission rate exceeding 2%. Even for sites with low utilization and high capacity gateway links, retransmission rates of 1%, and sometimes higher, were not uncommon.

Our results complement, extend, and bring up to date partial and incomplete results from some recent work, and show that TCP retransmissions continue to constitute a non-negligible percentage of the overall traffic, despite significant advances across the board in telecommunications technologies and network protocols. These results are in contrast to expectations that increased link capacities and improvements in TCP congestion control mechanisms may make retransmissions (and packet loss) negligible. Apart from their inherent value in quantifying salient aspects of the distributions for aggregate retransmission rates in the Internet, the results presented are pertinent to end-to-end protocol designers and evaluators; they provide a range of "realistic" scenarios under which, and a "marker" against which, simulation studies can be configured and calibrated, and protocols evaluated.